







\newcount\refnumber
\newcount\temp
\newcount\test
\newcount\tempone
\newcount\temptwo
\newcount\tempthr
\newcount\tempfor
\newcount\tempfiv
\newcount\testone
\newcount\testtwo
\newcount\testthr
\newcount\testfor
\newcount\testfiv
\newcount\itemnumber
\newcount\totalnumber
\refnumber=0
\itemnumber=0
\def\initreference#1{\totalnumber=#1
                 \advance \totalnumber by 1
                 \loop \advance \itemnumber by 1
                       \ifnum\itemnumber<\totalnumber
                        \temp=100 \advance\temp by \itemnumber
                        \count\temp=0 \repeat}

\def\ref#1{\temp=100 \advance\temp by #1
   \ifnum\count\temp=0
    \advance\refnumber by 1  \count\temp=\refnumber \fi
   \ [\the\count\temp]}

\def\reftwo#1#2{\tempone=100 \advance\tempone by #1
   \ifnum\count\tempone=0
   \advance\refnumber by 1  \count\tempone=\refnumber \fi
   \temptwo=100 \advance\temptwo by #2
   \ifnum\count\temptwo=0
   \advance\refnumber by 1  \count\temptwo=\refnumber \fi
 \testone=\count\tempone \testtwo=\count\temptwo
 \sorttwo\testone\testtwo
     \ [\the\testone,\the\testtwo]}       

\def\refthree#1#2#3{\tempone=100 \advance\tempone by #1
   \ifnum\count\tempone=0
    \advance\refnumber by 1  \count\tempone=\refnumber \fi
    \temptwo=100 \advance\temptwo by #2
   \ifnum\count\temptwo=0
    \advance\refnumber by 1  \count\temptwo=\refnumber \fi
    \tempthr=100 \advance\tempthr by #3
   \ifnum\count\tempthr=0
    \advance\refnumber by 1  \count\tempthr=\refnumber \fi
 \testone=\count\tempone \testtwo=\count\temptwo \testthr=\count\tempthr
 \sortthree\testone\testtwo\testthr
   \test=\testthr  \advance\test by -2
 \ifnum\test=\testone    \test=\testtwo  \advance\test by -1
    \ifnum\test=\testone   
    \ [\the\testone--\the\testthr]\fi \advance\temptwo by 1
  \else
     \ [\the\testone,\the\testtwo,\the\testthr]    
 \fi}

\def\reffour#1#2#3#4{\tempone=100 \advance\tempone by #1
   \ifnum\count\tempone=0
    \advance\refnumber by 1  \count\tempone=\refnumber \fi
    \temptwo=100 \advance\temptwo by #2
   \ifnum\count\temptwo=0
    \advance\refnumber by 1  \count\temptwo=\refnumber \fi
    \tempthr=100 \advance\tempthr by #3
   \ifnum\count\tempthr=0
    \advance\refnumber by 1  \count\tempthr=\refnumber \fi
    \tempfor=100 \advance\tempfor by #4
   \ifnum\count\tempfor=0
    \advance\refnumber by 1  \count\tempfor=\refnumber \fi
 \testone=\count\tempone \testtwo=\count\temptwo \testthr=\count\tempthr
 \testfor=\count\tempfor
 \sortfour\testone\testtwo\testthr\testfor
   \test=\testthr \advance\test by -1
   \ifnum\testtwo=\test   \test=\testtwo \advance\test by -1
    \ifnum\testone=\test  \test=\testfor \advance\test by -3
     \ifnum\testone=\test \ [\the\testone--\the\testfor]
     \else \ [\the\testone--\the\testthr,\the\testfor]
     \fi
    \else  \test=\testfor \advance\test by -1
     \ifnum\testthr=\test \ [\the\testone,\the\testtwo--\the\testfor]
     \else\ [\the\testone,\the\testtwo,\the\testthr,\the\testfor]
     \fi
    \fi
   \else \ [\the\testone,\the\testtwo,\the\testthr,\the\testfor]
   \fi}

\def\reffive#1#2#3#4#5{\tempone=100 \advance\tempone by #1
   \ifnum\count\tempone=0
    \advance\refnumber by 1  \count\tempone=\refnumber \fi
    \temptwo=100 \advance\temptwo by #2
   \ifnum\count\temptwo=0
    \advance\refnumber by 1  \count\temptwo=\refnumber \fi
    \tempthr=100 \advance\tempthr by #3
   \ifnum\count\tempthr=0
    \advance\refnumber by 1  \count\tempthr=\refnumber \fi
    \tempfor=100 \advance\tempfor by #4
   \ifnum\count\tempfor=0
    \advance\refnumber by 1  \count\tempfor=\refnumber \fi
    \tempfiv=100 \advance\tempfiv by #5
   \ifnum\count\tempfiv=0
    \advance\refnumber by 1  \count\tempfiv=\refnumber \fi
 \testone=\count\tempone \testtwo=\count\temptwo \testthr=\count\tempthr
 \testfor=\count\tempfor \testfiv=\count\tempfiv
 \sortfive\testone\testtwo\testthr\testfor\testfiv
  \test=\testthr \advance\test by -1
  \ifnum\testtwo=\test   \test=\testtwo \advance\test by -1
   \ifnum\testone=\test  \test=\testfor \advance\test by -3
    \ifnum\testone=\test \test=\testfiv \advance\test by -4
     \ifnum\testone=\test\ [\the\testone--\the\testfiv]
     \else\ [\the\testone--\the\testfor,\the\testfiv]
     \fi
    \else \ [\the\testone--\the\testthr,\the\testfor,\the\testfiv]
    \fi
   \else  \test=\testfor \advance\test by -1
    \ifnum\testthr=\test \test=\testfiv \advance\test by -2
     \ifnum\testthr=\test \ [\the\testone,\the\testtwo--\the\testfiv]
     \else \ [\the\testone,\the\testtwo--\the\testfor,\the\testfiv]
     \fi
    \else\ [\the\testone,\the\testtwo,\the\testthr,\the\testfor,\the\testfiv]
    \fi
   \fi
  \else \test=\testfor \advance\test by -1
   \ifnum\testthr=\test \test=\testfiv \advance\test by -2
    \ifnum\testthr=\test\
[\the\testone,\the\testtwo,\the\testthr--\the\testfiv]
    \else\ [\the\testone,\the\testtwo,\the\testthr,\the\testfor,\the\testfiv]
    \fi
   \else\ [\the\testone,\the\testtwo,\the\testthr,\the\testfor,\the\testfiv]
   \fi
  \fi}

\def\refitem#1#2{\temp=#1 \advance \temp by 100 \setbox\count\temp=\hbox{#2}}

\def\sortfive#1#2#3#4#5{\sortfour#1#2#3#4\relax
   \ifnum#5<#4\relax \test=#5\relax #5=#4\relax
     \ifnum\test<#3\relax #4=#3\relax
       \ifnum\test<#2\relax #3=#2\relax
         \ifnum\test<#1\relax  #2=#1\relax  #1=\test
         \else #2=\test \fi
       \else #3=\test \fi
     \else #4=\test \fi \fi}

\def\sortfour#1#2#3#4{\sortthree#1#2#3\relax
    \ifnum#4<#3\relax \test=#4\relax #4=#3\relax
       \ifnum\test<#2\relax #3=#2\relax
          \ifnum\test<#1\relax #2=#1\relax #1=\test
          \else #2=\test \fi
       \else #3=\test \fi \fi}

\def\sortthree#1#2#3{\sorttwo#1#2\relax
       \ifnum#3<#2\relax \test=#3\relax #3=#2\relax
          \ifnum\test<#1\relax #2=#1\relax #1=\test
          \else #2=\test \fi \fi}

\def\sorttwo#1#2{\ifnum#2<#1\relax \test=#2\relax #2=#1\relax #1=\test \fi}


\def\setref#1{\temp=100 \advance\temp by #1
   \ifnum\count\temp=0
    \advance\refnumber by 1  \count\temp=\refnumber \fi}

\def\printreference{\totalnumber=\refnumber
           \advance\totalnumber by 1
           \itemnumber=0
           \loop \advance\itemnumber by 1  
                 \ifnum\itemnumber<\totalnumber
                 \item{[\the\itemnumber]} \unhbox\itemnumber \repeat}


  \magnification=1200
  \hsize=15.5 truecm
  \vsize=23.0truecm
  \topskip=20pt            
  \fontdimen1\tenrm=0.0pt  
  \fontdimen2\tenrm=4.0pt  
  \fontdimen3\tenrm=7.0pt  
  \fontdimen4\tenrm=1.6pt  
  \fontdimen5\tenrm=4.3pt  
  \fontdimen6\tenrm=10.0pt 
  \fontdimen7\tenrm=2.0pt  
  \baselineskip=17.0pt plus 1.0pt minus 0.5pt  
  \lineskip=1pt plus 0pt minus 0pt             
  \lineskiplimit=1pt                           
  \parskip=2.5pt plus 5.0pt minus 0.5pt
  \parindent=15.0pt
\font\subsection=cmbx10 scaled\magstep1
\font\section=cmbx10 scaled\magstep 2
\font\bx=cmr8
\def\lsim{\; \raise0.3ex\hbox{$<$\kern-0.75em\raise-1.1ex\hbox{$\sim$}}\; }
\def\gsim{\; \raise0.3ex\hbox{$>$\kern-0.75em\raise-1.1ex\hbox{$\sim$}}\; }
\def\jump{\vskip 1truecm}
\def\GeV{\rm GeV}

\def\lphi{$\lambda\phi^4$}
\def\Re{{\rm Re}\hskip2pt}
\def\Im{{\rm Im}\hskip2pt}
\def\pdm{\partial_{\mu}}
\def\pum{\partial^{\mu}}
\def\pdn{\partial_{\nu}}
\def\pun{\partial^{\nu}}
\def\loota{{\scriptstyle\sqcap\kern-0.55em\hbox{$\scriptstyle\sqcup$}}}
\def\Loota{{\sqcap\kern-0.65em\hbox{$\sqcup$}}}
\def\box{\Loota}
\def\cL{{\cal L}}

\def\cO{{\cal O}}

\def\cL{{\cal L}}
\def\1{\aa}
\def\vek#1{{\rm\bf #1}}

\def\inv#1{\frac{1}{#1}}
\def\del{\partial}
\def\ra{\rangle}
\def\la{\langle}

\def\sign{{\rm\ sign}}
\def\pp{{\rm p}}
%
%
\def\dbk#1{\ \!\!\!{d\,^{#1}k\over (2\pi)^{{#1}}}}
\def\dbl#1{\ \!\!\!{d\,^{#1}l\over (2\pi)^{{#1}}}}
\def\sc{\sigma_c}
\def\Abs#1{\left|#1\right|}
\def\avek#1{\Abs{\vek{#1}}}
\def\wk{\omega_\vek{k}}
\def\wl{\omega_\vek{l}}
\def\wp{\omega_{\vek{p}+\vek{k}+\vek{l}}}

\def\SR{\Sigma_R(p_0,\vek{p};m)}
\def\SI{\Sigma_I(p_0,\vek{p};m)}
\def\etu{{\lambda^2\over 8(2\pi)^4}}

%

\def\np#1#2#3{{\it  Nucl.\ Phys.\ }{{\bf #1} {(#2)} {#3}}}
\def\pr#1#2#3{{\it Phys.\ Rev.\ }{{\bf #1} {(#2)} {#3}}}

\def\ptp#1#2#3{{\it  Prog.\ Theor.\ Phys.\ }{{\bf #1} {(#2)} {#3}}}

\def\rmp#1#2#3{{\it  Rev.\ Mod.\ Phys.\ }{{\bf #1}{(#2)}{#3}}}

\def\sovphjetp#1#2#3{{\it Sov.\ Phys.\ JETP\ }{{\bf #1} {(#2)} {#3}}}
\def\zp#1#2#3{{\it Z.\ Phys.\ } {{\bf #1}{(#2)}{#3}}}

\def\etal{{\sl et\ al. }}

\def\frac#1#2{{#1\over#2}}
\newcount\eqnumber
\eqnumber=1
\def\chaphead{}

\def\new{\hbox{(\chaphead\the\eqnumber}\global\advance\eqnumber by 1}
\def\eqref#1{\advance\eqnumber by -#1 (\chaphead\the\eqnumber
     \advance\eqnumber by #1 }
\def\first{\hbox{(\chaphead\the\eqnumber{a}}\global\advance\eqnumber by 1}
\def\last{\advance\eqnumber by -1 \hbox{(\chaphead\the\eqnumber}\advance
     \eqnumber by 1}
\def\eq#1{\advance\eqnumber by -#1 equation (\chaphead\the\eqnumber
     \advance\eqnumber by #1}
\def\eqnam#1{\xdef#1{\chaphead\the\eqnumber}}


\def\eqt#1{Eq.~({{#1}})}

\def\eqand#1#2{Eqs.~({$#1$}) \rm and ({$#2$})}
\centerline {\section 1. Introduction}
\jump
Phase transitions are often described in terms of an order parameter,
which in the electroweak theory is the background Higgs field. The states
with lowest energy are conveniently analyzed with the help of the effective
potential, which has lately been the subject of intensive studies in
connection with baryogenesis in the electroweak theory\ref{11}.
 It is however rather obvious that to investigate the dynamics of the order
parameter
in detail one should use the complete effective action, which includes also
terms depending on the derivatives of the background field. Although the ground
state has a constant value of the background field, locally there are
 fluctuations about this value, and they might turn out to be important
for local phenomena such as bubble nucleation in first order phase transitions.

To study the evolution of background fields one should solve the equation of
motion for the fluctuating background  as derived from the effective action,
including possible dissipation.  As the full effective action is an object that
is very difficult to estimate, one might resort to a  derivative expansion of
the effective action, cut it off after the first few terms and solve the
resulting equation of motion. This approach does not always work at finite
temperature   because the derivative expansion does not necessarily
exist\ref{8}
in the neighborhood of the origin of the  $(p_0,{\bf p})$--plane.

If the background field is assumed to be close to a constant value so that we
may consider just small space--time dependent perturbations about that constant
field, it is possible to expand the effective action in terms of the
perturbation  while retaining all the derivatives. In this approach one is able
to see not only the dispersion of the fluctuation but also which fields, or
modes of the field, are unstable and what are their actual decay rates. In what
follows we shall adopt this approach for $\lambda\phi^4$--theories.
We shall compute the effective action in this approximation up to two loops
and study the evolution and stability of the fluctuations, comparing eventually
the decay rates with the  Hubble rate to demonstrate how a realistic system
might actually behave.

The high temperature limit of the effective action of the
$\lambda\phi^4$--model has
recently been considered at one--loop level by Moss \etal\ref{1} (see
also\ref{20}). There are
several
reasons for reconsidering their work. First we want an independent
check of their results. They used an approach based on the
local-momentum space method for curved spacetime introduced by
Bunch and Parker\ref{2}, while we calculate it in a much simpler way through
Feynman diagrams.
Secondly, in\ref{1} the effective action was expanded
in powers of derivatives about a constant field. It is known\ref{8} that the
finite temperature correction
to the two-point function
is non-analytic at $p_\mu=0$, having different limits when, for instance,
$\{p_0=0,\vek{p}\rightarrow 0\}$ and $\{p_0\rightarrow 0,\vek{p}=0\}$.
It is, therefore, in general not meaningful to expand about the origin,
but only in fixed directions in the $(p_0,\vek{p})$--plane. These
expansions can be used to analyze e.g. space or time
independent solutions.
Recalling that the variation of the effective action gives an
equation of motion for $\la \phi(x)\ra$ we shall estimate the
Fourier spectrum of a solution to the equation of motion. As an
example we might consider fluctuations about the minimum of the
effective potential and therefore shift the field by a constant
to get to
the minimum. Then the deviation from the constant field
satisfies the on-shell condition $p_0^2=\vek{p}^2+m^2$, so
for a spatially almost constant field we could
expand around $\{p_0=m,\vek{p}=0\}$. Here we have neglected
corrections from interactions and quantum fluctuations but they have to be
small in order for perturbation theory to work. If we were interested
in other solutions than plane waves (such as solitons) we should
study the size of the derivatives anew.

A third reason for studying the \lphi--theory anew is
the existence of a two--loop diagram of order $\lambda^2$
which dominates over the one-loop diagram at high temperature
and when the constant part of the background field, $\sc$, is close to zero.
The one--loop tadpole in the $\lambda\phi^4$--theory,
depicted in Fig.\ 1, goes like $\lambda T^2$
for high $T$ but it is momentum independent so that the only
one--loop correction to the kinetic term comes from the
diagram in Fig.\ 2. It has a high $T$ behavior
like $\lambda^2\sc^2 T$. A naive estimate of the $T$ dependence
of the two--loop diagram in Fig.\ 3 is $\lambda^2 T^2$
(though we shall see in Sect.\ 3.2 that the mass
correction goes like $\lambda^2 T^2 \ln T$) so it dominates
for large enough $T$. Also, it turns out that an on-shell
imaginary part of the effective action, which is important
for studying the evolution of perturbations, first
arises at two loops.

The paper is organized as follows. In Section 2 we present the derivative
expansion and the generalized tadpole method for computing the
effective action. In
Section 3 we evaluate the effective action at one and two loop orders for
small non--constant fluctuations about a constant background. We also comment
on the existence of a derivative expansion of the effective action.
In Section 4
we discuss the dispersion relations and the decay rates for the fluctuating
modes. We show that although most fluctuations will decay during the course
of the cosmic expansion, some will still be present at the phase transition.
Section 5 contains our conclusions and a discussion of some open
problems.
\jump
%
\centerline{\section 2. Methods for computing effective action}
\jump
\centerline {\subsection 2.1 The derivative expansion}
\jump
The effective action $\Gamma [\phi]$ is the Legendre transformation of the
generating functional $W[J]$
\eqnam{\defW}$$
e^{iW[J]} = {\cal N}\int D\Phi e^{iS[\Phi] + i\int d^4xJ(x)\Phi(x)}\ ,
\eqno\new)$$
where $S[\Phi]$ is the tree level action of the quantum field $\Phi$, and
${\cal N}$ is a
normalization constant.
Thus the effective action of the classical background field $\phi$ is defined
by
\eqnam{\defgamma}$$
\Gamma[\phi] = W[J] -  \int d^4 x\, {\delta W[J]\over\delta J(x)} J(x)\ ,
\eqno\new)$$
where $J$ has been eliminated using the definition of classical field
$\phi(x) = {\delta W[J]\over\delta J(x)}$, assumed to be invertible. From a
calculational point of view it is often better to express the effective action
by an equivalent functional integral form
\eqnam{\functgamma}
$$
e^{i\Gamma[\phi]} = {\cal N}\int D\Phi e^{iS[\Phi + \phi] + i\int d^4x
J(\phi)\Phi}\ ,
\eqno\new)
$$
where now $J(\phi) = -{\delta \Gamma\over \delta \phi}$, as can be directly
verified from the definition of effective action.  One can then expand the
effective action $\Gamma[\phi]$ about an arbitrary
field $\sigma (x)$:
\eqnam{\expandgamma}
$$
\Gamma[\phi(x)]= \sum_{n = 0}^{\infty}{1 \over n!}\int d^4x_i
\Gamma^{(n)}[\sigma;x_i]\prod^n_i [\phi(x_i)-\sigma(x_i)]\ ,
\eqno\new)
$$
where the Taylor coefficients of the expansion are given by
\eqnam{\onepi}$$
\Gamma^{(n)}[\sigma;x_i] = {\delta^n\Gamma\over \delta \phi(x_1) \dots \delta
\phi(x_n)}\bigg |_{\phi = \sigma}\ ,
\eqno\new)$$
which are 1PI Green functions with a background field $\sigma$. In general
$\Gamma^{(n)}[\sigma;x_i]$ is a functional of $\sigma(x)$ and a non--local
function of $x_i$. If we want to write it as a local function we need to expand
in the derivatives of the fields as well.  We should however remember
that the effective action is genuinely non-local both in time and space (e.g.\
through logarithmic terms from loop corrections).

Let us assume for the  moment that the effective action can be written as a
derivative expansion of the classical field $\phi$, which is equivalent to the
statement that the Green functions are analytical at the origin, when expressed
in momentum space. Then
\eqnam{\deregamma}$$
\Gamma[\phi (x)] = \sum_{k=0}^{\infty}\int d^4x \Gamma_{2k}(\phi, \partial\phi,
\box\phi, \dots)\ ,
\eqno\new)$$
where each term $\Gamma_{2k}[\phi, \partial \phi, \box\phi,\dots]$ is of the
order $2k$ with respect the partial derivative operators. $\Gamma_2$ can always
be written in the unique form $\Gamma_2=\frac 12 Z[\phi]\partial_\mu
\phi\partial^\mu \phi$; all functionals with two derivatives can always be cast
into this form by partial integration\ref{20} (note that $\Gamma_2$ cannot in
general be written as  $\inv{2}Z(\phi)\phi\box\phi$). We may thus write down
the
effective action to second order in derivatives
by noting that by definition $Z$ and $V_{eff}\equiv-\Gamma^{(0)}[\phi]$ do not
depend on derivatives,
and thus their functional forms do not depend on whether $\phi$ is a constant
or not. The easiest way to compute them is therefore to consider a
{\it constant} background field $\sigma_c$ and afterwards replace
$\sigma_c\rightarrow\phi$ in $V_{eff}[\sigma_c]$ and $Z[\sigma_c]$.

It is possible to follow a similar procedure  for an arbitrary $k$ (and also
for fields other than scalars). One must
first find a basis, in which it is possible to express all the terms containing
$k$ derivatives in a unique fashion. For example, in the case $k = 4$, we
find that the basis is formed by the set $\{\pdm\phi\pum\phi\pdn\phi\pun\phi,\
\pdm\phi\pum\pun\phi\pdn\phi,\ \pdm\pdn\phi \pun\pum\phi\}$. Thus $\Gamma_4$ is
uniquely represented by the   linear combination
\eqnam{\g4terms}$$
\Gamma_4 = \Gamma_{4,1}[\phi]  \pdm\phi\pum\phi\pdn\phi\pun\phi +
\Gamma_{4,2}[\phi] \pdm\phi\pum\pun\phi\pdn\phi + \Gamma_{4,3}[\phi]
\pdm\pdn\phi \pun\pum\phi\ .
\eqno\new)$$

Clearly, different physical problems warrant different expansions. The
derivative expansion is suitable for e.g. slowly varying background fields (but
possibly with large amplitudes),
whereas  modes that approximately obey the on--shell
condition are best described in terms of an expansion in small perturbations.
%
%
\jump
\centerline{\subsection 2.2 The tadpole method}
\jump
A frequently used method for computing the effective potential
is to first calculate the tadpole of the theory where the
field $\phi$ has been shifted by a constant, $\sigma_c$, and then
integrate it with respect to $\sigma_c$\ref{7}. In this way an infinite
class of diagrams is resummed. The reason is that the mass gets
a contribution from shifting the field and that contribution
can be resummed to all orders using the effective $\sigma_c$--dependent
mass. We shall here extend the tadpole method to the whole effective action.

Suppose the effective action is expanded about some field $\sigma(x)$ (we
consider only a real scalar field for simplicity, but the method can easily be
extended to other fields).
Since $\Gamma [\phi(x)]$, given by \eqt{\expandgamma}, is independent of the
expansion point
$\sigma(x)$ we can take the functional derivative
of \eqt{\expandgamma} with respect
to $\sigma (y)$ to obtain
\eqnam{\dEA}
$$
\eqalign{
\sum_{n=0}^\infty \int d^4x_i & \Bigl\{
\frac{\delta \Gamma^{(n)} [\sigma(x);x_i]}
{\delta \sigma (y)} \prod_{i=1}^n (\phi(x_i)- \sigma(x_i))-
  \cr &
  \Gamma^{(n)} [\sigma(x);x_i] \sum_{j=1}^n
\delta(x_j-y) \prod_{i \neq j}(\phi(x_i)-\sigma(x_i)) \Bigr\} =0\ . \cr}
\eqno\new)
$$
If we  put $\phi(y)=\sigma(y)$ in \eqt{\dEA} and use  $\Gamma [\sigma (x)]=
\Gamma^{(0)} [\sigma (x)]$
we get
\eqnam{\tadpole}
$$
\frac{\delta\Gamma [\sigma(x)]}{\delta \sigma (y)} =
\Gamma^{(1)}[\sigma(x);y]\ ,
\eqno\new)
$$
which is the tadpole equation for the effective action.
The right hand side of \eqt{\tadpole} is just the tadpole for a
theory  where the field has been shifted by a non-constant field
$\sigma(x)$.

To see how the calculations work out in practise let us consider
the Lagrangian
\eqnam{\Lphi4}
$$
\cL = \inv{2} \left( \del_\mu\phi\del^\mu\phi-m^2\phi^2\right) -
\frac{\lambda}{4!}\phi^4\ .
\eqno\new)
$$
After shifting the field to $\phi(x)+\sigma(x)$ we get effectively
a space--time dependent mass term that cannot be dealt with exactly. This
means also that we cannot resum the same infinite set of diagrams
as for the effective potential. To proceed we have to consider a
perturbation series in the non-constant part of $\sigma (x)$.
Thus we write
\eqnam{\shift}
$$
\sigma(x)=\sigma_c+b(x),
\eqno\new)
$$
where $\sigma_c$ is
constant and $b(x)$ is small. Higher orders are not important when the
fluctuation energy is dominated by the non--interacting part or when
$b^2\lsim (\pp^2+m^2)/\lambda$, where at finite temperature one should replace
$m$ by the plasma mass $m(T)$. (When there is no risk of
confusion, we shall denote the Fourier transform of
$b(x)$ simply by $b(p)$).

The Lagrangian for the shifted field is
\eqnam{\shift}
$$
\eqalign{
\cL (\phi(x)+\sigma(x)) = & \cL(\sigma(x)) +
\frac{\delta \Gamma_{cl}}
{\delta\sigma(x)} \phi(x) + \inv{2} \left( (\del \phi(x))^2-
(m^2+\frac{\lambda^2\sigma_c}{2})\phi^2(x)\right) \cr
&-
\frac{\lambda}{2}
(\sigma_c b(x)+\frac{b^2(x)}{2})\phi^2(x)
 - \frac{\lambda(\sigma_c+b(x))}{3!}
\phi^3(x) -\frac{\lambda}{4!}\phi^4(x)\ , \cr}
\eqno\new)
$$
where $\Gamma_{cl}[\sigma(x)]$ is the classical action for the
Langrangian in \eqt{\Lphi4}. Because of the space
dependence in $b(x)$ there are some new Feynman rules that do
not conserve momentum. The rules are given in Fig.\ 4.

 Note that  the Fourier transform $(b^2)(q)$ is not equal to $(b(q))^2$ but
 is the Fourier
transform of $b(x)^2$.
If we can choose $\int d^4x b(x) = 0$, then $b(x)$ only includes
symmetric fluctuations around a constant field. Then we still have
$(b^2)(q=0)=\int d^4x b^2(x)>  0$ if $b(x)$ is not identically
zero but measures the average size of the fluctuation. In a diagram with all
the external momenta equal to zero
(i.e. for the effective potential) there still remains a contribution
from  $(b^2)(q=0)$.

Let us now take a look at the structure of the perturbative
expansion of the effective action. We can consider an expansion
in $\lambda$, $b$ and $\hbar$, but since we are interested in
quantum effects we set the loop expansion parameter $\hbar$  equal to
one. Higher loops are also
suppressed by higher
powers of $\lambda$. The tree-level tadpole
just gives the classical action. At one-loop and to $\cO (b)$
we have the diagrams in Fig.\ 5.
The first diagram can be integrated with respect to $\sigma_c$
and we get the usual one-loop effective potential $V_{eff}(\sc)$.
If we expand
the vertex function in $p$ in the second and third diagrams we
have to zeroth order in $p$ and first order in $b$
$$
\eqalign{&\cr
i\Gamma^{(1)}_{1-loop}[\sigma(x)]-&i\Gamma^{(1)}_{1-loop}[\sigma_c]\simeq\int
\dbk{4}\left(\inv{2}\frac{\lambda}{k^2\hskip-1pt -M^2}+
\inv{2}\frac{\lambda^2\sc^2}{(k^2\hskip-1pt -M^2)^2}\right) b(-p) \cr
& = \inv{2}\frac{\del}{\del\sc}\int\dbk{4}\frac{\lambda\sc}{k^2-M^2}
b(-p) =
-i\frac{\del^2}{\del\sc^2} V_{eff}(\sc) b(-p) \ ,\cr}\eqno\new)
$$
where\eqnam{\mass}
$$M^2=m^2+\lambda\sc^2/2\ .
\eqno\new)
$$
This is what we get if we first replace $\sc$ by $\sc+b(x)$
in $\del V_{eff}(\sc)/\del\sigma_c$
and then expand to first order in $b$. We can easily understand
this from the derivative expansion of the effective action, as was discussed in
Sect.\ 2.1.

The tadpole diagrams that are linear in $b(p)$ are, of course,
essentially the two-point functions for the theory with shifted
$\sc$. In the following we calculate instead $\Gamma^{(2)}$
directly and one can construct the
effective action from \eqt{\expandgamma}. The effective action in
momentum space to $\cO(b^2)$ is
\eqnam{\EA}
$$
\Gamma[\sc,b]=-V_{eff}(\sc)+\Gamma^{(1)}(\sc) b(k=0)
+\inv{2}\int\dbk{4}b(-k)\Gamma^{(2)}(\sc;k)b(k)\ ,
\eqno\new)
$$
and $\Gamma^{(1)}$ is zero at the minimum of $V_{eff}$.
\jump
%
\centerline{\section 3. Finite temperature
effective action}
\jump
There are at least two formalisms for calculating vertex
functions at finite temperature, the imaginary and real
time formalisms (ITF and RTF).
They should both give the same result when used correctly.
In  ITF there is a naive way of extracting the leading
high $T$ behavior by only including the mode with zero
Matsubara frequency (n=0) (for bosons). For the
diagrams under consideration here this method is not reliable.
The correct method should include
a summation over all Matsubara frequencies and an analytic continuation
at the end.
RTF is easy to use for one-loop diagrams but for the two-loop
diagrams we have employed instead ITF with the convenient method developed in
\refthree{12}{13}{14}.
%
\jump
\centerline{\subsection 3.1 One-loop diagrams}
\jump
Let us start by giving the well known result for the zero
temperature part. The effective potential can be found in\ref{17}
 and we repeat it here for completeness:
\eqnam{\VeffT01loop}
$$
V_{eff}(\sc)=\frac{\sc^2}{2}m^2+\frac{\sc^4}{4!}\lambda
+\frac{M^4}{64\pi^2}\ln\left(\frac{M^2}{m^2}\right)
+A+B\sc^2+C\sc^4\ .
\eqno\new)
$$
The renormalized two--point function is (see Fig. 2)
\eqnam{\1lact}
$$
\Gamma_{1-loop}^{(2)}[\sigma(x)]
=\frac{\lambda^2\sc^2}{32\pi^2}
\left[-2\sqrt{\frac{4M^2-p^2}{p^2}}\
\arctan\sqrt{\frac{p^2}{4M^2-p^2}} -
\ln\left(\frac{M^2}{4\pi\mu^2}\right)\right]
\eqno\new)
$$

$$
\simeq\left\lbrace
\eqalign{
\frac{\lambda^2\sc^2}{32\pi^2}&
\left[-2+\frac{p^2}{6M^2}-
\ln\left(\frac{M^2}{4\pi\mu^2}\right)\right]
+\cO(p^4)\ , \cr
\frac{\lambda^2\sc^2}{32\pi^2}&
\left[(2-\frac{\pi}{\sqrt{3}})-(\frac{2\pi}{3\sqrt{3}}-1)
\frac{p^2-M^2}{M^2}-
\ln\left(\frac{M^2}{4\pi\mu^2}\right)\right]
+\cO((p^2-M^2)^2)\ .\cr}\right .
 $$
While the expansion around $p^2=0$ agrees with\reftwo{3}{2}, we should
like to point out that the expansion
around $p^2-M^2$ is actually better suited for solutions close to the mass
shell.

The finite temperature effective potential has been calculated
in several papers (see e.g.\ref{16}) and we just give the result
(see Appendix A for the definition of $F^4_1$)
\eqnam{\EP1loop}
$$
V_{eff,\beta}(\sc)=-\inv{6\pi^2} F^4_1(T,M^2)
\simeq -\frac{T^4\pi^2}{90}+\frac{T^2m^2}{24}
+\frac{\sc^2}{2}\frac{\lambda T^2}{24}
-\frac{TM^3(\sc)}{12\pi}\ .
\eqno\new)
$$
The dominant mass correction at high $T$ is $\lambda T^2/24$.

Next we look at the effective action at finite $T$ and here
we notice some problems if we estimate the high $T$ limit
using only the zero Matsubara frequency mode. In that
approximation we would find\eqnam{\fint1}
$$
\eqalign{&\cr
\Gamma^{(2)}_{\beta,n=0}(p_0,\vek{p})=&
\frac{\lambda^2 \sc T}{16\pi}\inv{\pp}\left[
\arcsin\left(\frac{\pp^2+p_0^2}
{\pp\sqrt{4M^2+\pp^2-2p_0^2+p_0^4/\pp^2}}\right) \right.\cr &\cr
-&\left.
\arcsin\left(\frac{\pp^2-p_0^2}
{\pp\sqrt{4M^2+\pp^2-2p_0^2+p_0^4/\pp^2}}\right)
\right],
\cr &\cr}\eqno\new)
$$
where $\pp=|\vek p|$.
This function has a well defined limit when $p_\mu\rightarrow0$
and a threshold at $\{p_0=M,\vek{p}=0\}$. These properties
are however not shared by the correct two-point function. In particular
the threshold is unphysical since there is no decay process allowed
at that momentum\ref{15}. To get the correct answer we can use either
ITF or RTF to obtain\eqnam{\correct}
$$
\Gamma^{(2)}_\beta (p_0,\vek{p}) =
-\frac{\lambda^2\sc^2}{16\pi^2}
\int_0^\infty\frac{dk\,kf_B(\omega)}{\omega}
\inv{\pp}\ln \left(\frac{(p_0^2-\pp^2+
2\pp k)^2-4p_0^2\omega^2)}
{(p_0^2-\pp^2-2\pp k)^2-4p_0^2\omega^2)}\right)\ ,
\eqno\new)
$$
where
$$
\omega = \sqrt{k^2+M^2}\ , \ \ f_B(\omega) = \inv{e^{\beta\omega}-1} \ .
$$
If we set $p_0=0$, expand in small
$\avek{p}$ and take the high $T$ limit we get\eqnam{\hight}
$$
\eqalign{
\Gamma^{(2)}_\beta (0,\vek{p})=&
-\frac{\lambda^2\sc^2}{16\pi^2}
\int_0^\infty\frac{dk\,kf_B(\omega)}{\omega}
\frac{2}{\pp }\ln\Abs{\frac{\pp -2k}{\pp +2k}}
\cr
&\simeq
\frac{\lambda^2\sc^2}{16\pi}\frac{T}{M}(1-\frac{\pp^2}{12M^2})\ .
\cr}
\eqno\new)
$$
where the $\pp$--term coincides with the result
of\ref{1}.{\footnote{*}{\bx\baselineskip=9pt Note that there is a misprint in
Eq.\ (24) in\ref{1}.}
A possibility of avoiding the problem of non-analyticity at the origin
is to  expand around the
mass shell. When we expand
around $\{p_0^2=M^2,\vek{p}=0\}$ we find
\eqnam{\EAXXX}
$$
\Gamma^{(2)}_\beta (p_0,\vek{p})\simeq
\frac{\lambda^2\sc^2}{16\pi} \frac{T}{M}
\left(2(2-\sqrt{3})+\frac{7\sqrt{3}-12}{3}\  \frac{p_0^2-M^2}{M^2}
-\frac{2(2-\sqrt{3})}{3}\ \frac{\pp^2}{M^2}\right)\ .
\eqno\new)
$$
One can, of course, obtain higher derivative terms by simply
expanding to higher order in $(p_0^2-M^2)$ and $\pp^2$.

The expansion in \eqt{\EAXXX} could be used to write
down the dispersion relation near mass shell. We shall return to dispersion
relations in Sect.\ 4. After performing the resummation of the tadpole to get
the  thermal mass of $\lambda T^2/24$, no particularly
interesting feature is found in the high $T$ limit.
The only appreciable effect is the thermal mass. For instance, on--shell there
is no imaginary part at the one--loop level. Such an imaginary part, which is
of essential importance when one wants to study the evolution of the
perturbations, is
first generated at the two--loop level.
%
\jump
\centerline{\subsection 3.2 Two-loop diagrams}
\jump
For $T/M$  large, higher loop diagrams go as higher powers of $T$ than the
one--loop diagram. On the other hand they are also suppressed
by higher powers of $\lambda$. At two-loop order the diagrams in
Figs. 3 and 6 are of the order $\lambda^2$,
all other are of higher order.
The leading diagram in $T/M$ to each
order in $\lambda$ is the one with as many tadpole
insertions as possible. As can be seen from \eqt{\EP1loop}
it gives a $\lambda T^2/24$ contribution which can be
resummed by just shifting to an effective temperature
dependent mass $M^2+\lambda T^2/24$.
This is well known at the one-loop level\ref{16}
but at the two--loop level one must be careful not to
double count diagrams. Parts of the leading ``double-bubble" (Fig.\ 6) are
already included in the one loop tadpole when the mass
is shifted.
An easy way of keeping track of the counting is
to introduce a finite, $T$ dependent counter term as in\ref{9}.
In the ``setting sun" (Fig.\ 3), however, the leading
corrections can be resummed using a $T$-dependent
mass. In this Section we perform the calculation without
resumming tadpoles, and defer the resummation to Sect. 4.

The  diagram in Fig.\ 6 does not depend on the
momentum so it can be absorbed in the effective potential.
It has been calculated both at zero\ref{3} and
finite\ref{6} temperature. As for the one-loop
case we state the result for the effective potential
for completeness
\eqnam{\VT02loop}
$$
\eqalign{
V^{(2)}_{eff}(\sc) =
 \frac{\lambda m^2}{8(4\pi)^4}
&\left\{ \frac{\lambda\sc^2}{m^2}\left[ \ln^2\Bigl(\frac{M^2}{m^2}\Bigr)
-5\ln\Bigl(\frac{M^2}{m^2}\Bigr)\right] +\frac{M^4}{m^4}
\ln^2\Bigl(\frac{M^2}{m^2}\Bigr)\right\}\cr
&+A+B\sc^2+C\sc^4 \ .\cr  }
\eqno\new)
$$
At finite $T$ the result is \eqnam{\VTF2loop}
$$
\eqalign{& \cr
V^{(2)}_{eff,\beta}(\sc) = &
\frac{\lambda}{32\pi^4}
\left(F^2_1(T,M)\right)^2
+\frac{\lambda^2\sc^2M^2}{128\pi^4}I(M/T) \cr
&+
\frac{\lambda M^2}{128\pi^4}
  \left(\inv{2}+\ln\Bigl(\frac{M^2}{m^2}\Bigr)\right) F^2_1(T,M)\cr
&+
\frac{\lambda^2\sc^2}{128\pi^4}
\left( \bigl(\frac{\pi}{\sqrt{3}}-\inv{2}\bigr)
+\ln\Bigl(\frac{M^2}{m^2}\Bigr)
\right) F^2_1(T,M)
-\frac{\nu_1''(\sc)}{256\pi^4} F^2_1(T,M)\ .\cr &\cr}
\eqno\new)
$$
Here $\nu_1''(\sc)$ is a second order polynomial of $\sigma_c$ that
depends on the renormalization condition at one-loop,
and $F^2_1(T,M)$ and $I(M/T)$ are defined in Appendix A. We have not
verified the expressions in
\eqand{\VT02loop}{\VTF2loop},
which can be found in\reftwo{3}{6},
but instead concentrated on the high temperature expansion
of \eqt{\VTF2loop}. Most of it was derived in\ref{6}
but here we also show that (see Appendix A)
\eqnam{\ITexp}
$$
I(T,M) \simeq \frac{T^2}{M^2} \frac{5\pi^2}{24}
\ln\Bigl(\frac{M^2}{T^2}\Bigr) + \cO(T^2)\ .
\eqno\new)
$$
There is also a subleading term that goes like $T^2$ which
need not be numerically small compared to the one given
in \eqt{\ITexp}.

When it comes to the effective action
the finite temperature part of the ``setting sun" is particularly
interesting since it dominates over the one-loop ``bubble"
at high temperature $T\gg\sc^2/M$.
In the ``setting sun" there is a $T$ dependent
UV divergent
constant term but that cancels against the ``double-bubble"
diagram so that there are no $T$ dependent infinities when all
$\cO(\lambda^2)$ diagrams are included.

The derivative term
in the ``setting sun" is finite (since there is no $p$-dependent
term in the ``double-bubble" that can cancel infinities) and one might be
tempted to
try high $T$ expansion using  the zero mode
approximation. Expanding in $p_0$ and $\vek{p}$ one finds, after some algebra,
that\eqnam{\wrong}
$$
\eqalign{
\Gamma^{(2)}& = \frac{\lambda^2T^2}{6}\int\dbk{3}\dbl{3}
\inv{\vek{k}^2+M^2}\inv{(\vek{k}+\vek{l})^2+M^2}
\inv{-p_0^2+(\vek{p}-\vek{l})^2+M^2} \cr
& \simeq
 \frac{\lambda^2}{576\pi^2}\frac{T^2}{M^2}
\left(A+p_0^2-\inv{9}\pp^2\right)\ , \cr}
\eqno\new)
$$
where $A$ is an UV divergent momentum and temperature
independent constant.
Our experience from the one--loop calculation in Sect 3.1 shows, however, that
we cannot trust the zero mode approximation.
In fact, although the imaginary part of $\Gamma^{(2)}$ at
$\{p_0=M+i\epsilon,\vek{p}=0\}$ is UV finite, when it is
calculated from \eqt{\wrong} it is zero, whereas the correct value
is\ref{9}
$$
\Im \Gamma^{(2)}(p_0=M,\vek{p}=0)=\frac{\lambda^2T^2}{768\pi}\ .
\eqno\new)
$$
Also, the leading $T$ dependence of the ``setting sun"
goes like $T^2\ln T$ as we shall show below.
We must therefore sum over all Matsubara frequencies and the easiest
method to do that is the one described in
\refthree{12}{13}{14} and \ref{4}.

The two-point function can then be written as
\eqnam{\Pisarski}
$$
\Gamma^{(2)}_{2-loop}(p_0,\vek{p}) =
\frac{\lambda^2}{6}\int\dbk{3}\dbl{3}\int_0^\beta d\tau
e^{i\nu\tau}\Delta(-\tau,\vek{k})\Delta(-\tau,\vek{l})
\Delta(\tau,\vek{p}+\vek{k}+\vek{l})\ ,
\eqno\new)
$$
where\eqnam{\defn}
$$
\Delta(\tau,\vek{k})=\inv{2\wk}(\frac{e^{\beta\wk}}{e^{\beta\wk}-1}
e^{-\wk|\tau|} + \inv{e^{\beta\wk}-1}e^{\wk|\tau|})\ .
\eqno\new)
$$
After carrying out the $\tau$--integral  we find \eqnam{\result1}
$$
\eqalign{&\cr
\Gamma^{(2)}&_{2-loop}(p_0,\vek{p}) =
-\frac{\lambda^2}{8}
\int\dbk{3}\dbl{3}\frac{1}{\wk\wl}
 \inv{p_0^2-(\wk+\wl+\wp)^2}\cr
&-\frac{\lambda^2}{4}
\int\dbk{3}\dbl{3}\frac{1}{\wk\wl}f_B(\wk)
\sum_{\pm} \inv{(p_0\pm\wk)^2-(\wl+\wp)^2}\cr
&-\frac{\lambda^2}{8}
\int\dbk{3}\dbl{3}\frac{1}{\wk\wl}f_B(\wk)f_B(\wl)
\sum_{\pm,\pm} \inv{(p_0\pm\wk\pm\wl)^2-\wp^2}\ , \cr &\cr}
\eqno\new)
$$
where the summations are over the two and
four possible combinations of $\pm$.
We keep only the terms
with one and two distribution functions $f_B(\omega)$
since we are only interested in the finite temperature part.
The  term with one distribution function has a temperature
dependent, but momentum
independent, infinite part which cancels when all contributions
are added. We also subtract this part to get a finite answer.
All the angular integrations can then be performed explicitly
but we leave one integral for notational simplicity.
$\Gamma^{(2)}_{2-loop}(p_0,\vek{p})$ has also an
imaginary part which we determine through an analytical
continuation corresponding to the retarded two-point
vertex ($p_0\rightarrow p_0+i\epsilon$). The final result is\eqnam{\final}
$$
\eqalign{
\Gamma^{(2)}_{2-loop}(p_0,\vek{p}) =&-\frac{\lambda^2}{8(2\pi)^4}
\int_0^\infty \frac{dk\,dl\,kl}{\wk\wl}\Bigl [f_B(\wk)\bigl(
G_1^{(2)}(p_0,\vek{p})+\delta G_1^{(2)}(p_0,\vek{p})\bigr)\cr
&+f_B(\wk)f_B(\wl) G_2^{(2)}(p_0,\vek{p})\Bigr ]\ ,}
\eqno\new)
$$
where $\delta G_1^{(2)}(p_0,\vek{p})$ regulates the
infinity\ref{9}, and we have
separated the terms with one and two distribution functions:\eqnam{\finaldef}
$$
\eqalign{ &\cr
G_1^{(2)}(p_0,\vek{p})&=
 \inv{2\pp}\int_{\Abs{k-\pp}}^{k+\pp} dz
\sum_{\pm} \Biggl\{
\ln\Abs{\frac{p_0^2-(\pm\wk+\wl+\omega_{l-z})^2}
{p_0^2-(\pm\wk+\wl+\omega_{l+z})^2}}  \cr
&+
i\pi
\left[ \theta\left( p_0^2-(\pm\wk+\wl+\omega_{l-z})^2\right)
 -\theta\left( p_0^2-(\pm\wk+\wl+\omega_{l+z})^2
\right )\right ]\Biggr\}\ ,\cr
G_2^{(2)}(p_0,\vek{p})&=
\inv{2\pp}\int_{\Abs{k-\pp}}^{k+\pp} dz
\sum_{\pm,\pm} \Biggl\{
\ln\Abs{\frac{(p_0\pm\wk\pm\wl)^2-m^2-(z-l)^2}
{(p_0\pm\wk\pm\wl)^2-m^2-(z+l)^2}}  \cr
&+
i\pi \sign(p_0\pm\wk\pm\wl)
\bigl [ \theta\left ( (z-l)^2+m^2-(p_0\pm\wk\pm\wl)^2\right ) \cr
&-
\theta\left ( (z+l)^2+m^2-(p_0\pm\wk\pm\wl)^2
\right ) \bigr ]\Biggr\}\  ,\cr
\delta G_1^{(2)}(p_0,\vek{p})&=
 {2l\over \pp}\int_{\Abs{k-\pp}}^{k+\pp}\frac{dz}{z}\; .
   \cr
 &\cr}
\eqno\new)
$$
It is interesting to note that the two--loop result
\eqt{\final} is analytic at
the origin, unlike the one--loop action \eqt{\correct}.
The leading mass correction has been calculated explicitly
by Parwani\ref{9} at
high $T$ and at $\{p_0=M,\vek{p}=0\}$ (see Appendix A for details):
\eqnam{\T2lnT}
$$
\Gamma^{(2)}_{2-loop}(p_0=M,\vek{p}=0)
\simeq
-\frac{2}{3}\frac{\lambda^2}{128\pi^2} T^2\ln(\frac{M}{T})\ .
\eqno\new)
$$
One can now construct the effective action using \eqt{\EA}.
%
\jump
\centerline{\section 4. Evolution of perturbations}
\jump
When studying the evolution of perturbations about a constant field, the
knowledge about the two--point function is usually not enough. The action has,
apart from some special cases, also a linear part, so that
the equation of motion
is of the form $\Gamma^{(2)}(\sigma_c) b = - \Gamma ^{(1)}(\sigma_c)$. The
right hand side of this equation can be eliminated by
writing the field in the form
$b(x) = b_h(t) + \delta b(x)$. The first part is only time dependent
and satisfies
the equation of motion with vanishing spatial derivatives. It describes the
homogeneous rolling of the field on the effective potential surface
as long as $b_h(t)$ remains small. The second
part, $\delta b$, describes the fluctuations on the homogeneous field, and
satisfies
the dispersion relation.

As an application we shall now consider the evolution of local perturbations
$b(x)$ around the high $T$ minimum $\sigma_c=0$, which is the initial state
for the phase transition, and for which $\Gamma^{(1)} \equiv 0$. The one--loop
contribution to $\Gamma^{(2)}$,
\eqt{\correct}, vanishes, and the lowest order correction
is provided by the two--loop diagram of Fig.\ 3 as given by \eqt{\final}. For
small $b(x)$, when we may neglect terms higher than ${\cal O}(b^2)$ in the
expansion for the effective action, the evolution of different Fourier modes is
independent and
given by the dispersion relation
 \eqnam{\dispersio}
$$
p_0^2=\vek{p}^2 +m_R^2+\SR +i\SI\ ,
\eqno\new)
$$
with \eqnam{\sigmas}
$$\eqalign{
\SR=&-\Re \left (\Gamma^{(2)}(p_0,\vek{p};m)-\Gamma^{(2)}(m,0;m)\right ),\cr
\SI=&-\Im \Gamma^{(2)}(p_0,\vek{p};m)\ , \cr}
\eqno\new)
$$
where the renormalized mass $m_R$ has been introduced to absorb the order
${\cal O}(\lambda^2)$ mass correction, part of which comes from
$\Gamma^{(2)}(m,0;m)$. We have also indicated explicitly the mass--dependence
of the two--point function. We may write
\eqnam{\thmass}
$$
m_R^2=m^2+{\lambda T^2\over 24}+{\cal O}(\lambda^2)\equiv m(T)^2+{\cal
O}(\lambda^2)\ ,
\eqno\new)
$$
where $m(T)$ is the familiar one--loop plasma mass. In fact, to order ${\cal
O}(\lambda^2)$ we may simply replace
the mass parameter $m$ by $m(T)$ in $\Gamma^{(2)}(p_0,\vek{p};m)$. This amounts
to plasma resummation of the propagators in the loop diagrams. (As discussed
before, there is no
double counting in this case). Thus to lowest order we may write, denoting
$\Im p_0 = -\gamma /2$ and $\Re p_0=\omega$, the
dispersion relation as
\eqnam{\disp2}
$$
\eqalign{
\omega^2= &\, \pp^2+m(T)^2-{1\over 4}\gamma^2+\Sigma_R(\omega,
\vek{p};m(T))\ ,\cr
\gamma = & -\Sigma_I(\omega, \vek{p};m(T))/\omega\ . \cr}
\eqno\new)
$$

We may solve \eqt{\disp2} approximately by setting
$\omega^2\simeq\omega_{\pp}^2\equiv {\pp}^2+m(T)^2$. To this end we have
studied the behavior of $\SR$ and $\SI$
numerically. It turns out that when $\pp\rightarrow\infty$,
the contribution from the part with two distribution functions becomes
unimportant. Roughly, numerically we find that
$\gamma \sim T^3/\pp^2$ which is also an upper limit that can be derived
analytically. We also find that for large p the real part is independent
of p and goes roughly like $(2\pi)^{-4}\lambda^2\ln T$, so that
for $\lambda \ll 1$ this part is never important for the dispersion
relations. The limit $\pp\rightarrow\infty$ is thus completely dominated
by the part with one
distribution function, $G_1^{(2)}+\delta G_1^{(2)}$ in \eqt{\final}.
Scaling the relevant part of \eqt{\final} by $\pp$
and taking the limit, it is possible to show that (see Appendix B for details)
\eqnam{\limit}
$$\eqalign{
\Gamma^{(2)}(\omega_p,\vek{p}&)\simeq\etu F^2_1(T,m(T))\left [\ln {\pp^2\over
m(T)^2} + i\pi\right ],\cr
&^{|\vek{p}|\rightarrow\infty}\ \cr}
\eqno\new)
$$
where $F^2_1$ has been defined in Appendix A, and for large $T$ and
$\lambda\ll 1$, we have $F^2_1\sim \pi^2T^2/6$. In the
high temperature limit the dispersion
relation \eqt{\disp2} for the real part reads then
$p_0^2=\pp^2+m(T)^2+3(2\pi)^{-4}\lambda m(T)^2\ln (\pp^2/m(T)^2)$ so
that $\SR$ never contributes significantly.
Thus $\omega =\omega_{\pp}$ is a self--consistent approximate solution.

As discussed in\ref{15}, $\gamma$ is the rate by
which a given plane wave mode
thermalizes.
We have computed $\Im \Gamma^{(2)}$ from \eqt{\final} numerically, and the
resulting thermalization rate is drawn in Fig.\ 7 for a few temperatures. The
imaginary part exists
for all four--momenta obeying the (approximate) dispersion relation, and this
means that every small perturbation $b(x)$ about the high $T$ minimum
$\sigma_c=0$ will eventually smoothen out and thermalize.

The essential question then is, how rapid is the rate of thermalization as
compared with the expansion rate of the Universe, given by the Hubble
parameter $H$. Assuming the Standard Model degrees of freedom,
at high temperature we may write \eqnam{\hubble}
$$
H=\left ( {8\pi^3g_*\over 90}\right )^{1\over 2}{T^2\over M_P}\simeq
17{T^2\over M_P}\ ,
\eqno\new)
$$
where $M_P=1.22\times 10^{22} \GeV$ is the Planck mass. Comparing ({\limit})
and
({\hubble}) we find that all momenta obeying \eqnam{\obey}
$$
\pp\lsim 2.8\lambda^2\times 10^{14} \GeV\; ,
\eqno\new)
$$
decay in less than a Hubble time.

Recall that the critical temperature
and the intrinsic mass scale (the inverse of the wall thickness
of a bubble) are typically related by $\sqrt{\lambda}T_c\sim M$.
This means that perturbation modes of the size of the critical bubble have
not yet decayed by the onset of phase transition provided
$\lambda \lsim 10^{-5}(M/100\ \GeV)$. For such small $\lambda$ the inherent
local
perturbations would certainly be important for bubble nucleation.
%
\jump
\centerline{\section 5. Conclusions}
\jump
We have found that the finite temperature effective action
can be relatively easily computed using the tadpole method generalized
for space--time dependent fields in Sect.\ 2.2. In that case resummation is
however not
possible except for constant background fields $\sigma_c$.
Nevertheless,
one can do perturbation theory in small fluctuations $b(x)$ about a constant
background, which in effect amounts to computing Green functions for the
shifted theory. To lowest non--trivial order one then just evaluates the
two--point function, which we did to two--loop order in Sect.\ 3. After that
it is straightforward to write down dispersion relations and to study the
evolution of the fluctuations. Since this only requires an expansion of the
effective action about the mass shell we do not encounter the problem
of non--analyticity at $p_\mu=0$. We also observe that the zero
Matsubara frequency approximation for the leading high temperature
terms is not correct in general. In our case, even the
UV--finite imaginary part needs a summation over all Matsubara
frequencies.

Our main conclusion is that small space--time dependent perturbations about the
high temperature minimum do not automatically thermalize before the phase
transition starts. This would naturally affect bubble formation in a
first order phase transition, although to what extent is not clear. In spite of
the fact that this conclusion holds strictly speaking only
for the \lphi--model, we could view the \lphi--model as an effective theory
of the order parameter of a more realistic theory, such as the Standard Model.
The gross features of the present study would then presumably show up also
in the full Standard Model.

It is however clear that the estimate at the end of
Sect. 4 is very rough
and has to be improved for an application to a real
physical system.  The main issue is
the strength of the effective coupling. Some recent estimates
seem to indicate that the temperature dependence can
be quite strong, and that the four--point interaction for zero momentum
particles goes in fact to zero at the critical temperature of a second order
phase transition\reftwo{18}{19}. This is caused by the vanishing mass
at the critical temperature. The coupling constant decreases at the
same time as the mass and thus the relevant p also decreases.
It is not clear what the net effect of the coupling constant renormalization
is. Furthermore, the electroweak phase transition is likely to be
first order, albeit possibly very weakly so. These questions can only
be addressed after a more detailed study of the damping rates in
the full electroweak theory\ref{10}.

We have only considered small perturbations. It is conceivable that large
perturbations could, because of the non--linear coupling of the modes,
contain instabilities so that some perturbation modes would actually
get amplified. Further studies in this direction are certainly  needed.

\vfill\eject
%
\centerline{\subsection Appendix A}
\jump

In this Appendix we give some useful expansions employed in Sect.\ 3 and
calculate the leading $T$ dependence
of \eqt{\final} as well as the function $I(T/M)$ in
\eqt{\VTF2loop}.

The function
$$
F^m_n(T,M)=\int_0^\infty \frac{dk\,k^m}{\omega^n} f_B(\omega)
\ ,\ \ \ \omega=\sqrt{k^2+M^2}\ ,\eqno(A1)
$$
often occurs in loop calculation results
at finite temperature (cf Eqs. ({\VTF2loop}), ({\limit})).
It satisfies the
relation
$$
\frac{\del F^m_n}{\del M^2} = -\frac{m-1}{2} F^{m-2}_n\ ,
\ \ m>1\ .\eqno(A2)
$$
In this paper we need $F^2_1$ and $F^4_1$ and they have the high
$T$ expansions
$$
\eqalign{
F^2_1(T,M) &= \frac{\pi^2 T^2}{6}-\frac{\pi TM}{2} -
\frac{M^2}{4}\ln(\frac{M}{4\pi T}) + \ldots \cr
F^4_1(T,M) &= \frac{\pi^4T^4}{15}-\frac{\pi^2 T^2 M^2}{4} +
\frac{\pi TM^3}{2} +\frac{3M^4}{16}\ln(\frac{M}{4\pi T})
+\ldots \cr}\eqno(A3)
$$
\smallskip
\smallskip
Let us now turn to the leading $T$ dependence
of \eqt{\final} and the function $I(T/M)$ in
\eqt{\VTF2loop}.
The leading high $T$ expansion of \eqt{\final}
was derived in\ref{9}. We have checked the coefficients
of the $T^2\ln T$ terms in the part with two distribution functions, using a
different method, and
found them to be correct.

The essential integral in the part
with two distribution functions of \eqt{\final}
is
$$
T^2I_1(T/M) = \int_0^\infty\frac{dk\,dl\,kl}{\wk\wl}
f_B(\wk)f_B(\wl)\ln\Abs{\frac{k-l}{k+l}}\ ,\eqno(A4)
$$
where  $ a=M/T$ and $\{p_0=M,\vek{p}=0\}$.
We expect $I_1$ to go like
$\ln a$ for small $a$ so we compute the derivative
with respect to $a^2$.  Using
$$
\frac{d}{da^2}\left(\frac{f(\omega_x)}{\omega_x}
\frac{f(\omega_y)}{\omega_y}\right) =
\frac{f(\omega_y)}{\omega_y}\inv{2x}\frac{d}{dx}
\left(\frac{f(\omega_x)}{\omega_x}\right)+
\frac{f(\omega_x)}{\omega_x}\inv{2y}\frac{d}{dy}
\left(\frac{f(\omega_y)}{\omega_y}\right)\ ,\eqno(A5)
$$
where $\omega_x=\sqrt{x^2+a^2}$ and $f(\omega_x)=(e^{\omega_x}-1)^{-1}$,
we find after partial integration
$$
 \frac{dI_1}{da^2}=\int_0^\infty\frac{dx\,dy}{\omega_x\omega_y}
f(\omega_x)f(\omega_y) \simeq \inv{a^2}\frac{\pi^2}{4}\
$$
for small $a$. After integration with respect to $a^2$ we get
$$
T^2I_1(T/M)=-\frac{\pi^2}{2}T^2\ln(\frac{T}{M})+\cO(T^2)\ .\eqno(A6)
$$

The function $I(T/M)$ in \eqt{\VTF2loop} can be
expanded in a similar manner. It is defined as\ref{6}
$$
I(T/M) = \frac{T^2}{M^2} \int_0^\infty
\frac{dx\,dy\,xy}{\omega_x\omega_y}
f(\omega_x)f(\omega_y)\ln
\Abs{\frac{(a^2+2xy)^2-4\omega_x^2\omega_y^2}
{(a^2-2xy)^2-4\omega_x^2\omega_y^2}}\ .\eqno(A7)
$$
The same trick works and we find
$$
I(T/M)= -\frac{5}{6}\frac{\pi^2}{2}
\frac{T^2}{M^2}\ln(\frac{T}{M})+\cO(T^2)\ ,\eqno(A8)
$$
where the factor of 5/6 comes from
$$
\int_0^\infty\frac{dx\,dy}{(x^2+1)(y^2+1)}
\frac{2x^4-x^2y^2+2y^4}{(x^2-xy+y^2)(x^2+xy+y^2)}
= \frac{5}{6}\frac{\pi^2}{2}\ .\eqno(A9)
$$
It is the same diagram, the ``setting sun", that
leads to $I$ and $I_1$ but $I$ is computed at
$p_0=0$ since it occurs in the effective potential
while $I_1$ is computed on-shell, $p_0=M$.
We therefore do not expect them to be equal but
only of the same order of magnitude and to have
the same sign.
%
\jump\centerline{\subsection Appendix B}\jump
The asymptotic behavior of $\Gamma^{(2)}$ can be calculated from \eqt{\final}
for all temperatures in terms of the functions $F^m_n$. In the effective
action the part with one distribution function,
$G^{(2)}_1+\delta G^{(2)}_1$ in \eqt{\final},
 dominates over $G^{(2)}_2$, and thus the behavior of the two--point function
in the limit $\pp\rightarrow \infty$ is solely determined by
$G^{(2)}_1+\delta G^{(2)}_1$. We
define
$$
I_2 = \int_0^\infty \hskip-1pt\frac{dk\,dl\,k}{\wk}f_B(\wk)
\int_{\Abs{k-\pp}}^{k+\pp} \hskip-3pt dz \inv{2\pp}
 \sum_{\pm} \left\{ \frac l\wl
\ln{p_0^2+\hskip-1pt i\hskip-1pt\epsilon -(\pm\wk+\wl+\omega_{l-z})^2 \over
p_0^2+\hskip-1pt i\hskip-1pt\epsilon -(\pm\wk+\wl+\omega_{l+z})^2} + \frac
{4z}\wl\hskip-1pt \right \}.
\eqno (B1)
$$
$I_2$ can be evaluated in the limit $\pp \rightarrow \infty$. After scaling the
integration variables $k,\ l$ and $z$ by $\pp$ we can extract the leading
behavior of the integral $I_2$. The logarithmic piece has now a finite,
non-zero
limiting value, and to the leading order we can take this limit in the
integrand. We obtain
$$ \eqalign{
I_2 \simeq &\int_0^\infty \hskip-1pt\frac{dk\,dl\,k}{\omega_{pk}}\pp^3
f_B(\omega_{pk})\cr
&\times\int_{\Abs{1 - k}}^{1 + k}\hskip-2pt dz \frac 12
 \sum_{\pm} \left\{
\ln{1\hskip-1pt + \hskip-1pt i\epsilon -(l +|l-z|\pm k )^2 \over
1 \hskip-1pt +\hskip-1pt i\epsilon - (2l + z\pm k )^2} + \frac {4z}{\sqrt{l^2 +
 {M^2}/{\pp^2}}} \right \}\ .\cr}
\eqno (B2)
$$
Note that we cannot take the limit in the whole integral, because  both the $k$
and $l$--integrations are singular at the origin when $\pp \rightarrow \infty$.
The easiest way to proceed  is to first perform the $l$ -integration and after
that the $z$--integration. This yields
$$
I_2 \simeq - \int_0^\infty dk {\pp^3 k\over \sqrt{\pp^2 k^2 + M^2}}
f_B(\omega_{pk})\left [ 2k\ln{\pp^2\over M^2} + i2\pi k\right]\ ,
\eqno (B3)
$$
which, apart from prefactors, is $\Gamma^{(2)}$ in \eqt{\final}.
By rescaling $k \rightarrow k/\pp$ we obtain the result  Eq.\ (\limit).
The logarithmic behavior of the real part of the two point function is due to
the IR--singularity in the UV--regulating term.

\vfill \eject\null
\jump
\centerline {\subsection References}
\jump
\refitem{1}
      {I.~Moss, D.~Toms and A.~Wright,
      {\sl ``Effective action at finite temperature"},
      \pr{D46}{1992}{1671.}}

\refitem{2}
      {T.~S.~Bunch and L.~Parker,
      {\sl ``Feynman propagators in curved space-time:
      A momentum space representation"},
      \pr{D20}{1979}{2499.}}

\refitem{3}
      {J.~Iliopoulos, C.~Itzykson and A.~Martin,
      {\sl ``Functional methods and perturbation theory"},
      \rmp{47}{1975}{165.}}

\refitem{4}
      {R.~D.~Pisarski,
      {\sl ``Computing finite-temperature loops with ease"},
      \np{B309}{1988}{476.}}

\refitem{6}
      {Y.~Fujimoto, R.~Grigjanis and R.~Kobes,
      {\sl ``Two-loop finite temperature effective potential"},
      \ptp{73}{1985}{434.}}

\refitem{7}
      {S.~Y.~Lee and A.~M.~Sciaccaluga,
      {\sl ``Evaluation of higher-order effective
      potential with dimensional regularization"},
      \np{B96}{1975}{435.}}

\refitem{8}
      {A.~Weldon,
      {\sl ``Mishaps with Feynman parametrization
      at finite temperature"},
      \pr{D47}{1993}{594.}}

\refitem{9}
        {R.~R.~Parwani,
        {\sl ``Resummation in a hot scalar field theory"},
        \pr{D45}{1992}{4695.}}

\refitem{10}
      {P.~Elmfors, K.~Enqvist and I.~Vilja, work in progress.}

\refitem{11}
      { For a recent review see A.~G.~Cohen, D.~B.~Kaplan and A.~E.~Nelson,
      {\sl ``Progress in electroweak baryogenesis"},
      to appear in {\it Annual Review of Nuclear and
      Particle Science } {\bf 43}, and references therein.}

\refitem{12}
      {R.~Balian and C.~De~Dominicis,
      {\sl ``Sur la fonction de Green \`{a} une particule
      en m\'{e}canique statistique quantique"},
      \np{16}{1960}{502.}}

\refitem{13}
      {G.~Baym and A.~M.~Sessler,
      {\sl ``Perturbation-theory rules for computing the
      self-energy operator in quantum statistical mechanics"},
      \pr{131}{1963}{2345.}}

\refitem{14}
      {I.~E.~Dzyaloshinski,
      {\sl ``A diagram technique for evaluating transport coefficients
      in statistical physics at finite temperatures"},
      \sovphjetp{15}{1962}{778.}}

\refitem{15}
      {A.~Weldon, ``{\sl Simple rules for discontinuities
      in finite-temperature field theory~}",
      \pr{D28}{1983}{2007.}}

\refitem{16}
      {L.~Dolan and R.~Jackiw,
      ``{\sl Symmetry behavior at finite temperature~}",
      \pr{D9}{1974}{3320.}}

\refitem{17}
      {S.~Coleman and E.~Weinberg,
      {\sl ``Radiative corrections as the origin of spontaneous
      symmetry breaking"}
      \pr{D7}{1973}{1888.}}

\refitem{18}
        {P.~Elmfors,
        {\sl ``Finite temperature renormalization for the
       $(\phi^3)_6$-
        and $(\phi^4)_4$-models at zero momentum"},
        \zp{C56}{1992}{601.}}

\refitem{19}
      {N.~Tetradis and C.~Wetterich, {\sl ``The high temperature phase
      transition for $\phi^4$-theories"}, preprint DESY 92-093.  }

\refitem{20}
      {C.~M.~Fraser,
      {\sl ``Calculation of higher derivative terms in
      the one-loop effective Lagrangian"},
      \zp{C28}{1985}{101.}}
\printreference

\vfill\eject
\def\jumpp{\jump\noindent}
\centerline{\subsection Figure Captions}\jumpp
{\bf Fig.\ 1} The one--loop tadpole.\jumpp
{\bf Fig.\ 2} One--loop diagram for $\Gamma^{(2)}$.\jumpp
{\bf Fig.\ 3} Two--loop diagram for $\Gamma^{(2)}$ ("setting sun").
\jumpp
{\bf Fig.\ 4} Feynman rules for momentum--dependent external legs.
\jumpp
{\bf Fig.\ 5} One--loop diagrams for the effective action in the generalized
tadpole method.
\jumpp
{\bf Fig.\ 6} Two--loop tadpole ("double bubble").
\jumpp
{\bf Fig.\ 7} The thermalization rate $\gamma$ for on--shell configurations
$\omega\simeq\omega_\pp$ for $T=10\; m(T)$ (dash-dotted curve),\
$100\; m(T)$ (dotted curve) and $1000\; m(T)$ (solid curve).
\vfill\eject
\null
\nopagenumbers
\vskip -20pt
\line {\hfill NORDITA--93/22  P}
\line {\hfill hep-ph/9303269}
\vskip 1.2truecm
\centerline {\section Finite temperature effective action }
\vskip0.2truecm
\centerline{\section and thermalization of perturbations }
\vskip 1.5truecm
\centerline {Per Elmfors{\footnote{\bx $^\dagger$}{\baselineskip=9pt\bx
internet: elmfors@nordita.dk}}, Kari Enqvist{\footnote{\bx
$^\ddagger$}{\baselineskip=9pt\bx
internet: enqvist@nbivax.nbi.dk}}
and Iiro Vilja{\footnote{\bx $^\star$}{\baselineskip=9pt\bx
internet: iiro@nordita.dk}}}
\vskip 0.4truecm
\centerline {Nordita, Blegdamsvej 17, DK-2100 Copenhagen \O , Denmark}
\vskip 0.4truecm
\vskip 0.4truecm
\centerline {\bf Abstract}
\vskip 1truecm
\noindent
The effective action is computed  for the \lphi--theory at finite temperature
for small perturbations about a constant  background field,
using a generalized tadpole  method. We find the
complete effective action, including the real and
imaginary parts, to all orders in derivatives and to order
${\cal O}(\lambda^2)$.
We demonstrate that the high $T$ approximation, where only the zero
Matsubara frequency is included, is incorrect for the
imaginary part even though it is UV-finite.
The solutions of the dispersion relations show that initial
perturbations do not
necessarily thermalize fast enough to be absent at the onset of phase
transition.
\vfill\eject

\end